\documentclass{PoS}

\usepackage{cite}
\usepackage{graphicx}
\usepackage{amssymb,amsfonts,amsmath,bbold,latexsym}

\title{QCD-electroweak effects and a new prediction for Higgs production in gluon fusion process}

\ShortTitle{QCD/EW effects in gg $\to$ H}

\author{\speaker{Radja Boughezal}
\\
        Institut f\"ur Theoretische Physik,
        Universit\"at Z\"urich,\\
        Winterthurerstr. 190,
        8057 Z\"urich, Switzerland\\
        E-mail: \email{radja@physik.uzh.ch}}

\abstract{We discuss the recent derivation of the three-loop 
${\cal O}(\alpha\alpha_s)$ contribution to the Higgs boson production 
cross section via gluon fusion arising from diagrams with light quarks, 
using an effective theory approach. 
We show results for the updated prediction of this process accounting for
all the new theoretical calculations and the newest MSTW PDFs.}

\FullConference{European Physical Society Europhysics Conference on High
  Energy Physics, EPS-HEP 2009,\\
		 July 16 - 22 2009\\
		 Krakow, Poland}

\begin{document}
\section{Introduction}
\label{sec:introduction}
The search for the Higgs boson
is a primary goal of the CERN Large Hadron Collider
(LHC), and is a central part of Fermilab's Tevatron program.
Recently, the Tevatron collaborations reported a 95\% confidence level 
exclusion of a Standard Model Higgs boson with a mass in the range
$160-170 \,{\rm GeV}$~\cite{:2009pt}. 
The dominant production mode at both the Tevatron and the LHC, gluon fusion 
through top-quark loops, receives important QCD radiative 
corrections~\cite{Dawson:1990zj,Djouadi:1991tka,Spira:1995rr}.  
The inclusive result increases by a factor of 2 at the
LHC and 3.5 at the Tevatron
when perturbative QCD effects through next-to-next-to-leading order (NNLO) are 
taken into account~\cite{Harlander:2002wh}.
The theoretical uncertainty from effects beyond NNLO is estimated  to be
about $\pm 10\%$ by varying renormalization and factorization scales. 
At this level of precision, electroweak corrections to the Higgs signal
become important. A subset of diagrams, where the Higgs couples
to the W and Z bosons which subsequently couple to light quarks, was
investigated in~\cite{Aglietti:2004nj,Aglietti:2006yd}. These terms are not
suppressed by light-quark Yukawa couplings, and receive a multiplicity
enhancement from summing over the quarks.  A careful study of the full 2-loop
electroweak effects was performed in Ref.~\cite{Actis:2008ug}.  They increase
the leading-order cross section by up to $5-6\%$ for relevant Higgs masses.
An important question is whether these light-quark contributions receive 
the same QCD enhancement as the top quark loops. If they do, then the full 
NNLO QCD result is shifted by $+5-6\%$ from these electroweak corrections.  
If not, this $5-6\%$ increase from light quarks would 
be reduced to $1-2\%$ of the NNLO result.  
As this effect on the central value of the production cross section 
and therefore on the exclusion limits and future measurements is 
non-negligible, it is important to quantify it.  The exact computation
of the mixed electroweak/QCD effects needed to do so requires 3-loop 
diagrams with many kinematic scales, and 2-loop diagrams with four external 
legs for the real-radiation terms.  Such a computation is prohibitively 
difficult with current computational techniques.\\
In Ref.~\cite{Anastasiou:2008tj}, the QCD corrections to the light-quark terms 
in the Higgs production cross section via gluon fusion were computed using 
an effective theory approach.  This approach was rigorously justified
by applying a hard-mass expansion procedure to the full 3-loop corrections. 
In addition to that, the most up-to-date QCD prediction 
for the Higgs boson production cross section in this channel was provided 
for use in setting Tevatron exclusion limits. In this contribution, we sketch
the calculational approach and the results of this investigation 
discussed in detail in~\cite{Anastasiou:2008tj}.
\section{Calculational approach}
\label{sec:calc}
The cross section for Higgs boson production in hadronic collisions can be written as
\begin{eqnarray}
\sigma(s,M_H^2) &=& \sum_{i,j} \int_{0}^1 dx_1  \int_{0}^1 dx_2 \,f_{i/h_1}(x_1,\mu_F^2)  f_{j/h_2}(x_2,\mu_F^2) \int_{0}^1 dz\,\delta\left(z-\frac{M_H^2}{x_1 x_2 s}\right) \nonumber \\ &\times& z \,\hat{\sigma}_{ij}\left(z; \alpha_s(\mu_R^2), \alpha_{EW},M_H^2/\mu_R^2; M_H^2/\mu_F^2\right).
\end{eqnarray}
Here, $\sqrt{s}$ is the center-of-mass energy of the hadronic collision, $\mu_R$ and $\mu_F$ respectively denote the renormalization and factorization scales, and the $f_{i/h}$ denote the parton densities.  The quantity $z \hat{\sigma}$ is the partonic cross section for the process $ij \to H+X$ with $i,j = g,q,\bar{q}$.  As indicated, it admits a joint perturbative expansion in the strong and electroweak couplings.
Considering QCD and electroweak corrections and suppressing the scale
dependence for simplicity, the partonic cross section can be written as:
\begin{equation}
\hat{\sigma}_{ij} = \sigma^{(0)}_{\rm EW} \,G^{(0)}_{ij}\left(z\right)+\sigma^{(0)} \sum_{n=1}^{\infty} \left(\frac{\alpha_s}{\pi}\right)^n G^{(n)}_{ij}(z)
\label{pcsec2}
\end{equation}
%
%
%
%
%
\\
%
%
The QCD corrections to the one-loop diagrams coupling the Higgs boson 
to gluons via a top-quark loop are given by 
\[ G_{ij}(z;\alpha_s) = \sum_{n=1}^{\infty} \left(\frac{\alpha_s}{\pi}\right)^n G^{(n)}_{ij}(z)
\]

\begin{figure}
 \begin{center}
   \includegraphics[width=0.40\textwidth]{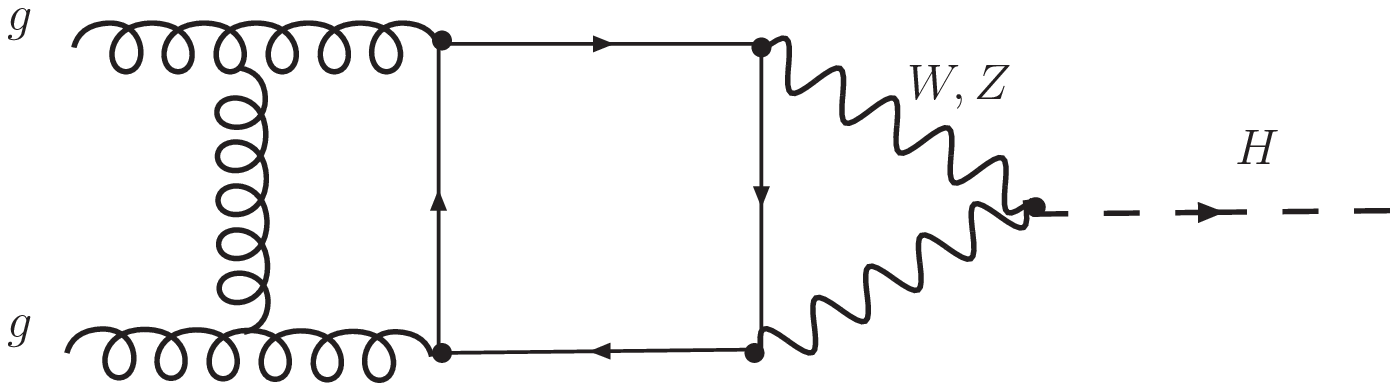}
   \includegraphics[width=0.40\textwidth]{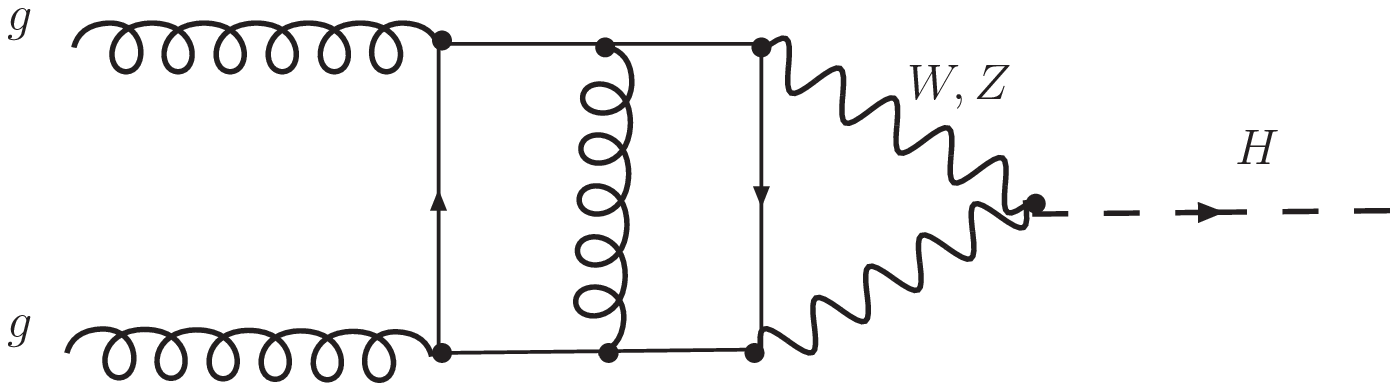}
   \caption{\textsf{Example three-loop light-quark diagrams contributing
            to the $C_{1w}$ term in the Wilson coefficient.}}
   \label{fig:3loop}
 \end{center}
\end{figure}

The cross section in Eq.~(\ref{pcsec2}) includes corrections to the
leading-order result valid through ${\cal O}(\alpha)$ in the electroweak
couplings and to ${\cal O}(\alpha_s^2)$ in the QCD coupling constant in the
large top-mass limit upon inclusion of the known results for $G^{(1,2)}_{ij}$.
Since the perturbative corrections to the leading-order result are large, it
is important to quantify the effect of the QCD corrections on the light-quark
electroweak contributions.  This would require knowledge of the mixed 
${\cal O}(\alpha\alpha_s)$. In lieu of such a calculation, the authors of
Ref.~\cite{Actis:2008ug} studied two assumptions for the effect of QCD
corrections on the 2-loop light-quark diagrams.
\begin{itemize}
\item {\it Partial factorization}: no QCD corrections to the light-quark
  electroweak diagrams are included. With this assumption, electroweak 
diagrams contribute only a $+1-2\%$ increase to the
Higgs boson production cross section.

\item{\it Complete factorization}: the QCD corrections to the electroweak
  contributions are assumed to be identical to those affecting the heavy-quark
  diagrams.
\end{itemize}
%
%
%
In this case the light-quark diagrams increase the full NNLO QCD production
cross section by $+5-6\%$. The last assumption was used in an earlier exclusion
of a SM Higgs boson of $170$ GeV by the Tevatron collaborations.
The calculation of the ${\cal O}(\alpha\alpha_s)$, which allows to check
these assumptions, can be done in the framework of an effective field theory 
where the W-boson is integrated out
\begin{equation}
{\cal L}_{eff} = -\alpha_s\frac{C_1}{4v} H G_{\mu\nu}^a G^{a\mu\nu}.
\label{lag}
\end{equation}
The Wilson coefficient $C_1$ arising from integrating out the heavy quark 
and the W-boson is defined through 
\begin{eqnarray}
C_1 &=& -\frac{1}{3\pi}\left\{1+\lambda_{EW}\left[1+a_s C_{1w}+a_s^2 C_{2w}\right]+a_sC_{1q} + a_s^2 C_{2q}  \right\}, \nonumber \\
C_{1q} &=& \frac{11}{4},\;\;\; C_{2q} = \frac{2777}{288} +\frac{19}{16}L_t+N_F
\left(-\frac{67}{96}+\frac{1}{3}L_t\right), \nonumber
\\
\lambda_{EW} &=& \frac{3\alpha}{16\pi
  s_W^2}\left\{\frac{2}{c_W^2}\left[\frac{5}{4}-\frac{7}{3}s_W^2+\frac{22}{9}s_W^4\right]+4\right\}, \nonumber
\end{eqnarray}
where $a_s = \alpha_s/\pi$, $N_F=5$ is the number of active quark flavors,
$L_t = {\rm ln}(\mu_R^2/m_t^2)$ and $s_W,c_W$ are respectively the sine and
cosine of the weak-mixing angle.
The Wilson coefficient obtained from using the complete factorization
assumption is given by
\[ C_1^{fac} = -\frac{1}{3\pi}\left(1+\lambda_{EW}\right) \left\{1+a_sC_{1q} + a_s^2 C_{2q}\right\} .\]
Factorization holds if $C_{1w}=C_{1q}$ and $C_{2w}=C_{2q}$. To test 
this assumption, the $C_{1W}$ coefficient was calculated 
in~\cite{Anastasiou:2008tj} by expanding the 3-loop 
QCD corrections to the light-quark electroweak diagrams, keeping the leading
term of that. The numerical effect of various choices for $C_{2w}$ was 
also studied. In Fig.~(\ref{fig:3loop}), sample diagrams involved in this 
calculation are shown.
\section{Results}
After a computation following the approach outlined above, we obtain the
following  result for $C_{1w}$:
\begin{equation}
C_{1w} = \frac{7}{6}.
\end{equation}
Two points should be noted regarding the comparison of this with 
the factorization hypothesis $C_{1w}^{fac} = C_{1q} = 11/4$.  
First, there is a fairly large 
violation of the factorization result:~$(C_{1q}-C_{1w})/C_{1w} \approx 1.4$.  
However, both expressions have the same sign, and a large difference from 
the $+5-6\%$ shift found before does not occur. 
%
%
In table~(\ref{tab:TEVxsec08}), the numerical results for the new prediction 
of the gluon fusion cross
section including all currently computed perturbative effects on the cross
section, are shown. These are: the NNLO $K$-factor
computed in the large-$m_t$ limit and normalized to the exact $m_t$-dependent LO
 result, the full light-quark electroweak correction and the ${\cal O}(\alpha_s)
$ correction to this encoded in $C_{1w}$, the bottom-quark contributions
using their NLO K-factors with the exact dependence on the bottom and top 
quark masses and finally the newest MSTW PDFs from
$2008$~\cite{Martin:2009iq}.
The new numerical values are $4-6\%$ lower than the numbers in 
Ref.~\cite{Catani:2003zt} used in an earlier exclusion of a SM Higgs boson 
mass of $170$ GeV with $95\%$ CL.
\begin{table}[t]
  \begin{center}
    \begin{tabular}{|c|c||c|c|}
      \hline
      $m_{H}$[GeV] &$\sigma^{best}$[pb]&$m_{H}$[GeV] &$\sigma^{best}$[pb]\\
        \hline \hline
     110&    1.417 ($\pm 7\%$ pdf) &160&0.4344 ($\pm 9\%$ pdf)
 \\[1mm]
      \hline
  115&    1.243 ($\pm 7\%$ pdf) &165&0.3854 ($\pm 9\%$ pdf)
\\[1mm]
      \hline
  120&    1.094  ($\pm 7\%$ pdf)  &170&0.3444 ($\pm 10\%$ pdf)
\\[1mm]
      \hline
  125 &   0.9669 ($\pm 7\%$ pdf) &175&0.3097 ($\pm 10\%$ pdf)
\\[1mm]
      \hline
  130 &  0.8570 ($\pm 8\%$ pdf) &180&0.2788 ($\pm 10\%$ pdf)
\\[1mm]
      \hline
  135 &   0.7620 ($\pm 8\%$ pdf) &185&0.2510 ($\pm 10\%$ pdf)
\\[1mm]
      \hline
  140 &   0.6794 ($\pm 8\%$ pdf) &190&0.2266 ($\pm 11\%$ pdf)
\\[1mm]
      \hline
  145 &   0.6073 ($\pm 8\%$ pdf) &195&0.2057 ($\pm 11\%$ pdf)
\\[1mm]
      \hline
  150 &   0.5439 ($\pm 9\%$ pdf) &200&0.1874 ($\pm 11\%$ pdf)
  \\[1mm]
      \hline
  155 & 0.4876  ($\pm 9\%$ pdf) & $-$& $-$
      \\\hline
    \end{tabular}
  \end{center}
  \caption{{Higgs production cross
   section (MSTW08) for Higgs mass values relevant for Tevatron, with
   $\mu= \mu_R =\mu_F=M_H/2 $.
   $\sigma^{best} =
   \sigma_{QCD}^{NNLO}+\sigma_{EW}^{NNLO}$~\cite{Anastasiou:2008tj}.  The
   theoretical errors PDFs are shown in the Table; the scale variation is
   $^{+7\%}_{-11\%}$, roughly constant as a function of Higgs boson mass.}
    \label{tab:TEVxsec08} }
\end{table}
\section{Conclusions}
In this contribution, we have briefly sketched the calculation of the mixed
QCD-electroweak corrections to the Higgs boson
production cross section in the gluon-fusion channel, due to diagrams 
containing light quarks. 
The leading term of this contribution was derived based on an effective 
Lagrangian obtained by integrating out the W-boson. This result allows 
us to check the factorization of electroweak and QCD
corrections proposed in Ref.~\cite{Aglietti:2006yd,Actis:2008ug}. 
Despite a fairly large violation of the factorization hypothesis, 
a significant numerical difference from the prediction of this hypothesis 
is not observed due to the structure of the QCD corrections. A new prediction
for the Higgs production cross section via gluon fusion was also presented.
The new numerical values are $4-6\%$ lower than the numbers in 
Ref.~\cite{Catani:2003zt} used in an earlier exclusion of a SM Higgs boson 
mass of $170$ GeV with $95\%$ CL.
\section{Acknowledgments}
This work is supported by the Swiss National Science Foundation 
under contract 200020-116756/2.

\end{document}